\documentclass{appolb}
\usepackage{graphicx}
\usepackage{bm,amsmath}

\begin{document}
\title{A numerical technique for preserving the topology of polymer
  knots\footnote{Proceedings of the 25th Marian Smoluchowski Symposium
  on Statistical Physics, 9--13 September 2012, Cracow. }
\thanks{The support of the Polish National Center of Science,
scientific project No. N~N202~326240, is gratefully acknowledged.
The simulations reported in this work were performed in part using the HPC
cluster HAL9000 of the Computing Centre of the Faculty of Mathematics
and Physics at the University of Szczecin.}}
\author{Yani Zhao and Franco Ferrari
\address{CASA* and Institute of Physics, University of Szczecin, Szczecin, Poland}} 
\maketitle
\begin{abstract}
The statistical mechanics of single polymer knots is studied using
Monte Carlo simulations. The polymers are considered on a cubic
lattice and their conformations are randomly changed with the help of
pivot transformations. After each transformation, it is checked if the
topology of the knot is preserved by means of a method called pivot
algorithm and excluded area (in short PAEA) and described in a
previous publication of the authors.
As an application of this method the specific energy, the radius of gyration and heat
capacity of a few types of knots are computed. 
The case of attractive short-range forces is investigated.
The  sampling of the energy states is performed by means of the
Wang-Landau algorithm.
The obtained results show that
the specific energy and heat capacity increase with increasing knot complexity. 
\end{abstract}
\section{Introduction}
Polymer knots are researched in connection with several applications,
mainly in biology and biochemistry
\cite{grosberg,dna,katritch96,katritch97,krasnow,
  laurie,cieplak,liu,wasserman,sumners}. 
In this work we investigate the statistical mechanics of a
single polymer knot by computing its specific energy, heat capacity
and gyration radius at different temperatures.
One of the main problems in studies of this kind is how to preserve
the polymer topology during the thermal fluctuations.
To this purpose, several methods have been
developed
\cite{vologodski,orlandini,kurt,mehran,pp2,metzler,muthu}. Most of
them are
based on the Alexander or Jones polynomials, which are rather powerful
topological invariants able to distinguish with a very high degree of accuracy the
different topological configurations. The main drawback
of these 
polynomials is that their calculation is time
consuming from the computational point of view.
In a recent work \cite{YZFF}, a strategy based on an excluded area
method called PAEA has been proposed in order to circumvent these difficulties.
A similar idea has been presented in \cite{swetnametal}, but in that
case, instead of using arbitrary pivot transformations to produce
random knot configurations, a set of topology preserving pull moves is
adopted. Within our method, instead, any transformation is
allowed. In
this way the equilibration of the system is faster and the access to
all possible conformations is easier.
Those transformations that lead to a change 
of topology are automatically detected by the PAEA algorithm and
discarded. 

According to \cite{YZFF}, one starts  
from a seed knot, for instance those given in
ref.~\cite{scharein}. Next, the knot
is changed at a randomly chosen set of segments by applying the pivot algorithm \cite{pivot}.
After each pivot move, it is easy to realize that the difference
between the old and new configurations, obtained by canceling the
segments that have been unaffected by the transformation, consists of
a closed loop. Around this loop, an arbitrary surface is stretched, whose
boundary is the closed loop itself. The criterion to reject changes that destroy the
topology of the knot is the presence or not of lines of the old knot that
cross such surface. If these lines are
present, the trial pivot move is rejected, otherwise is accepted. This combination of 
pivot algorithm and excluded area (PAEA) provides an efficient
and fast way to preserve the topology that can be applied to any
knot configuration, independently of its complexity. For pivot moves
involving a small number of segments the method becomes exact.
This technique may be employed in the study of the thermal and mechanical
properties of polymer knots as it has been done in Ref.~\cite{YZFF}. 
In this article we will extend that work by studying the case of
an attractive short-range potential, which is nothing else but a rough approximation
of the Lennard-Jones potential on the lattice.
Moreover, with respect to \cite{YZFF}, we do not limit ourselves to the
computation of the internal energy and heat capacity, but we consider
also the gyration radius of the knot. The
cases of the unknot, trefoil and $5_1$ topologies is investigated.
The sampling
of the canonical ensemble is achieved by using the Wang-Landau
algorithm \cite{wanglaudau} at different temperatures.
\\  
 
\section{Sampling and calculation method}
The way of generating random knot transformations
with the help of pivot transformations and the PAEA method needed to
prevent topology changes after these transformations have been already
extensively described in \cite {pivot} and \cite{YZFF}
respectively. We refer the interested reader to those publications for
more details.
In this Section we concentrate on the Wang-Landau (WL) method applied to
polymer knots. See also \cite{pnring} for applications
of this method to linear polymer chains and rings.

The WL algorithm can be regarded as a self-adjusting procedure for
obtaining the density of states 
$\Phi_i$:
\begin{equation}
 \int dX\delta(E_i-E(X))\equiv \Phi_i
\end{equation}
where $X$ is a microstate of the system under consideration. We
suppose here that the energy values are discrete, so that they can be
labeled by indexes $i,i',\ldots$. If the $\Phi_i$'s are known, then
the partition function can be constructed:
$Z=\sum_ie^{-\beta{E_i}}\Phi_i$\footnote{ 
We have put here
$\beta=T^{-1}$, where $\beta$ is the usual Boltzmann factor in
thermodynamic units in which the Boltzmann constant 
is equal to $1$.}.
As well, it is possible to
derive in an easy way the averages of any quantity that can be
expressed in terms of the momenta of the energy
$\langle E^l\rangle(\beta)=\sum_i E_i^l
e^{-\beta E_i}
\Phi_i/Z$, $l=1,2,\ldots$.
For example, the heat
capacity is given by:
\begin{equation}
 C(T) =\beta^2(\langle E^{2}\rangle(\beta) -(\langle
 E\rangle(\beta))^2)\label{heatcapacity} 
\end{equation}
According to the Wang-Landau method, the density of
states is constructed with successive approximations.
First, the would be density of states $g(E_i)$ is set to be
equal for all $E_i$'s by putting $g(E_i)=1$.
After that, a Markov chain of microstates $X_1,X_2,\ldots $ is
generated. The probability of transition from a state $X_i$ of energy
$E_i$ to a state $X_{i'}$ with energy $E_{i'}$ is
\begin{equation}
p(i\rightarrow
i')=\min[1,\dfrac{g(E_{i})}{g(E_{i'})}]\label{probabilitytrans} 
\end{equation}
If $p(i\rightarrow i')\ge 1$, the state $i'$ is automatically
accepted. If instead $p(i\rightarrow i')< 1$, then a random
number $0< \eta < 1$ is generated and the state is accepted only
if $p(i\rightarrow i')\ge \eta$.
Once an energy state $E_i$ is
visited, its corresponding would be 
density of states is updated by multiplying it by a modification factor $f$,
i.~e. 
\begin{equation}
g(E_i)\rightarrow fg(E_i)\label{fg}
\end{equation}
where 
$f>1$.
Moreover, the energy histogram  $H(E_{i})$
is updated by performing the replacement $H(E_{i})\rightarrow
H(E_{i})+1$~\cite{wanglaudau}. In this way, an energy state occurring
$N(E_i)$ times during the 
sampling will have a would be density of states $g(E_i)=f^{N(E_i)}$
and the transition probability 
(\ref{probabilitytrans}) to that state will be suppressed
by the factor $f^{-N(E_i)}$.
When the
energy histogram $H(E_i)$ becomes flat, 
then $g(E_i)$ converges to the density of states $\Phi_i$.
To show that, let's consider the probability ${\cal P}(E_i)$ of
obtaining a microstate $X$ with energy $E(X)=E_i$. This probability must be
equal to the probability of generating the state $X$ times the
probability of acceptance of $E_i$
introduced by the Wang-Landau algorithm, which is proportional to
$f^{-N(E_i)}$. In formulas:
\begin{equation}
{\cal P}(E_i)\propto f^{-N(E_i)}\Phi_i\label{proprel}
\end{equation}
The last factor in the above equation is due to the fact that the
probability of obtaining a microstate $X$ with energy $E_i$ is given by
$\frac{\int dX\delta(E_i-E(X))}{\sum_i\int
  dX\delta(E_i-E(X))}=\frac{\Phi_i}{\sum_i\Phi_i}$ and 
the denominator $\sum_i\Phi_i$ is an irrelevant constant. When the
energy histogram 
$H(E_i)$ becomes flat, this means that the
probability ${\cal 
  P}(E_i)$ is the same for all states $E_i$. In other words:
$
{\cal P}(E_i)=a$ for every $i$, where $a$ is a constant. Thus, from
Eq.~(\ref{proprel}) we obtain:
\begin{equation}
 f^{N(E_i)}=g(E_i)\propto\Phi_i
\end{equation}
Actually, if $f$ is too
big, the statistical errors on the $g(E_i)$'s may
grow large and the above equation is satisfied very roughly. 
On the other side, if $f$ is too small, it is necessary an enormous number of
microstates during the sampling in order to derive the $g(E_i)$'s.
For
this reason, in the Wang-Landau method the density of states is computed 
perturbatively. In the next step, one takes the $g(E_i)$ evaluated
with the modification factor $f$ as a starting point and generates
another Markov chain of microstates.
The energy histogram $H(E_i)$ is reset to $0$ and
all the previously explained procedure is repeated
for the new microstates apart from the fact that in Eq.~(\ref{fg})
$f$ is replaced by $\sqrt{f}$.
When the energy histogram becomes flat, the second approximation of
the $g(E_i)$'s is obtained. In the next approximations successive
square roots of $f$ are entering in the algorithm  until we arrive at
a step $n$ such that
$\sqrt[n]{f}=f_{final}\sim\exp(10^{-8})$~\cite{wanglaudau}.
The initial parameter $f$ is chosen in such a way that the simulations
will not take too much time and the statistical
errors on the $g(E_i)$ will not be too large.
%
%

In the present article the states are distinguished by the number $m$ of
closest contacts between the monomers, where $m$ takes
positive integer values. The meaning of contact in the present
context is explained in Refs.~\cite{YZFF,pnring}.
Short-range attractive forces are studied, so that
the
energy values are given by $E_{m}=m\varepsilon$, where  
$\varepsilon$ is the contact energy of two
unbonded monomers, which is negative in the attractive case.
Since the number of  samples is huge for
polymer systems, it is more convenient to consider the
logarithm of the density of states $\Omega_{m}=\ln
g(E_{m})$. In this way the transition probability is expressed as:
 \begin{equation}
p(m\rightarrow m') =\min[1,\exp(\Omega_m-\Omega_{m'})]
\end{equation}
and the modification factor becomes
$\ln(f)$. 
After  a state $E_m$ is visited, the corresponding energy histogram
should be updated by 
$H(E_{m})\rightarrow H(E_{m})+1$ and the density of state is modified by
$\Omega_m\rightarrow \Omega_m+ln(f)$. 

\section{Thermal properties of  polymer knots}
In this Section the thermal properties 
of a few types of polymer knots are studied.
In particular, the specific internal energy of the
polymer per unit of 
length $\langle E\rangle(\beta)/L$, the heat
capacity and the radius of gyration are computed in the case
of the unknot $0_1$, the trefoil
$3_1$ and the knot $5_1$. 

The gyration radius is not directly related to the moments of the
energy as mentioned in the previous Section. However, this quantity
may be computed by noticing \cite{pngyrrad} that
the mean square radius of gyration $\langle 
R_G^2
\rangle(\beta)$
can be written as follows:
\begin{equation}
\langle R_{G}^{2}
\rangle (\beta)  =\dfrac{\sum_{m}{\langle
    R_{G}^{2}\rangle}_{m}  e^{-\beta m 
    \varepsilon}\Omega_{m}}{\sum_{m} e^{-\beta m
    \varepsilon}\Omega_{m}} 
\label{rg}
\end{equation}
Here $\langle R_{G}^{2}\rangle_{m}=\langle \frac 1{L^2}\sum_{I,J=1}^{L}
\langle(\boldsymbol R_I-\boldsymbol R_J)^{2}
\rangle_m$ denotes the average of the gyration radius
computed over states with $m$ contacts. Moreover, $\boldsymbol R_{I}$ is the
position vector of $I$-th segment and $L$ is the length of the
polymer.

\begin{figure}
\centering
\includegraphics[width=5.3in]{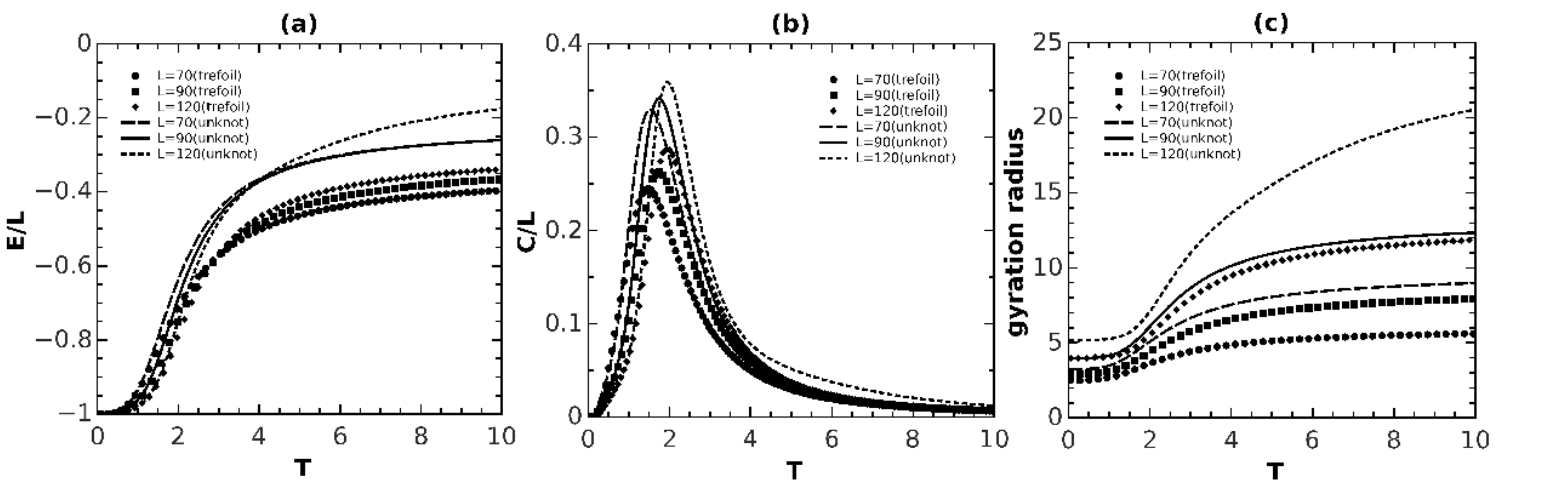}
\caption{\label{t-u}  Specific energy, specific heat capacity and
  gyration radius for the
  unknot and the trefoil as functions of the temperature. 
  Polymers with lengths  $L=70, 90$ and $120$ are considered. (a) Plot of the
  specific energy (in 
  $\varepsilon$-units); (b) Plot of the
  specific heat capacity (in $\varepsilon$-units); (c) Plot of the
  radius of gyration.}
\end{figure}
 
\begin{figure}
\centering
\includegraphics[width=5.3in]{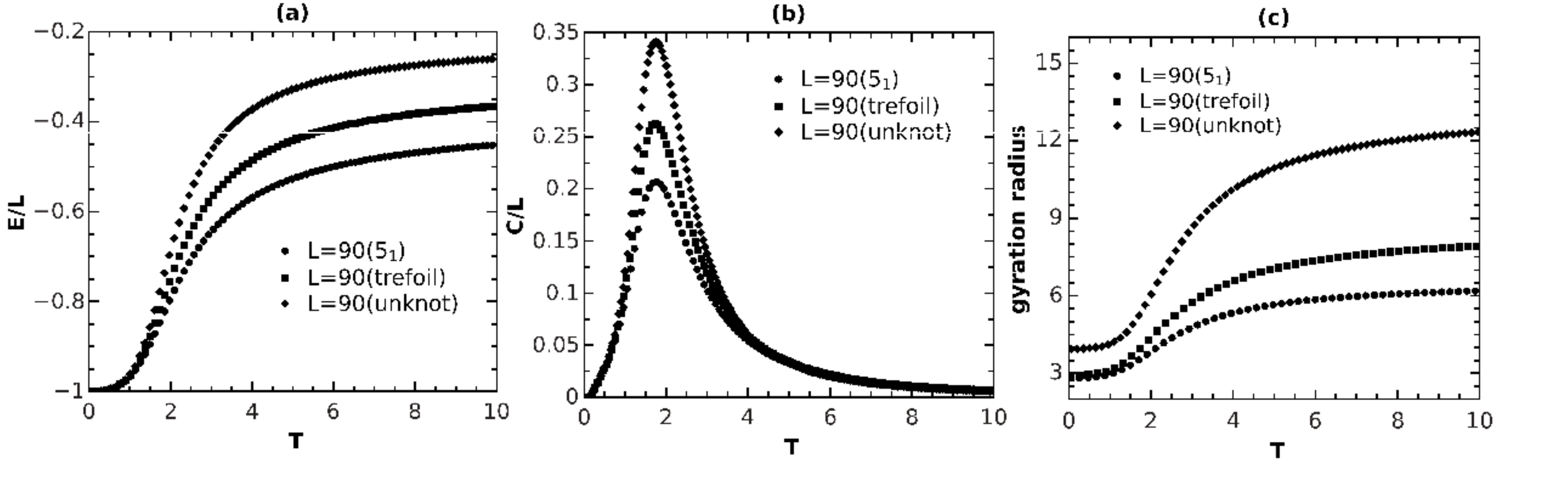}
\caption{\label{ut590}  Specific energy and heat capacities 
as functions of the
  temperature for the
  unknot, the trefoil and the knot $5_1$ 
  with length $L=90$. (a) Plot of the
  specific energy (in 
  $\varepsilon$-units); (b)
  Plot of the   specific heat capacity (in $\varepsilon$-units); (c)
  Plot of the gyration radius.}   
\end{figure}

In Fig.~\ref{t-u} are displayed the results for the unknot and the trefoil.
It is found that the growth of the specific 
energies $\langle E(\beta)\rangle/L$ in  
Fig.~\ref{t-u}(a) is characterised by three regions. At very low temperatures
the energy growth is practically zero because the
temperatures are too low to allow contacts between the monomers. When
the energy is enough to excite more states, the
specific energy grows 
rapidly as a 
function of the temperature until saturation is reached and the energy
increase becomes moderate. The fast increasing energy region causes a
peak of the heat capacity as shown in Fig.~\ref{t-u}(b) as it has been
explained in details in \cite{YZFF}. Concerning the
topological effects, we observe that both the energy and the heat
capacity grow with growing knot complexity, as it turns out from
Fig.~\ref{ut590} by comparing the plots for various knots with the same
length $L=90$. 
Fig.~\ref{t-u}(c) shows that in
the attractive energy case the mean square gyration radius
$\langle R_{G}^{2}
\rangle(\beta)$ grows with
increasing temperatures. This is an expected behavior.
As a matter of fact, in the case of attractive forces
the polymer turns out to be in a crumpled
conformation at very low temperatures 
with many contacts in order to minimize the energy.
On the contrary, at high temperatures the energy of the thermal
fluctuations is large with respect to $\varepsilon$, so that
the attractive forces become negligible.
Thus, the number of contacts $m$ will be in the average smaller at
higher
temperatures than at
lower energies.
As it is intuitive, a
smaller number of contacts $m$ corresponds to a larger volume occupied
by the knot,
which causes the observed increase of the radius of gyration with growing 
temperatures.  

\section{Conclusions}

We have studied the thermal properties of a few types of
polymer knots using the PAEA algorithm of Ref.~\cite{YZFF}. The
number of 
polymer segments affected by the pivot moves has been limited to
four. In this case the 
PAEA method is able to preserve the topology of the knot exactly \cite{YZFF}.
The results of \cite{YZFF}, which took into account only short-range
repulsive forces, have been extended to
attractive
forces and the calculation of the gyration radius has been added. 
The thermal properties of the unknot, the
knot $3_1$ and $5_1$ have been analysed with the 
help of Monte Carlo simulations based on the Wang-Landau algorithm and
the pivot method. 
A brief account about the Wang-Landau method has been provided.
The results, including those with the new topology $5_1$, confirm
those of Ref.~\cite{YZFF} even when the interactions are attractive. In
particular, the presence of the three 
regimes of  growth of the specific energy mentioned in the previous
Section has been observed.
Moreover, the role of the topological effects, which make
both the energy and the heat
capacity increase with increasing knot complexity, is confirmed.
The data coming from the calculation of the gyration radius give a
measure of the size of the polymer knot at different
temperatures. When the temperature is low, the number of contacts is at
its maximum and the gyration radius is at its minimum. The size of the
polymer points out to a possible crumpled conformation. At high
temperatures, the influence of the attractive forces becomes
negligible and the gyration radius attains slowly its maximum.
Even if we limited ourselves to
simple short-range 
interactions, there are no obstacles to extend our
procedure to  more realistic polymer systems.



\begin{thebibliography}{99}

\bibitem{grosberg} A. Yu. Grosberg Phys.-Usp. {\bf 40},  12  (1997).
\bibitem{dna} W. R. Taylor, Nature (London) {\bf 406}, 916 (2000).
\bibitem{katritch96} V. Katritch, J. Bednar, D. Michoud, R. G.  Scharein, 
J. Dubochet, A. Stasiak, Nature {\bf 384}, 142 (1996). 
\bibitem{katritch97} V. Katritch, W. K.  Olson, P. Pieranski, J. Dubochet 
and A. Stasiak, Nature {\bf 388}, 148 (1997). 
\bibitem{krasnow} M. A. Krasnow, A. Stasiak, S. J. Spengler, F. Dean, 
T. Koller and N. R. Cozzarelli, Nature {\bf 304}, 559 (1983).
\bibitem{laurie} B. Laurie, V. Katritch, J. Dubochet and A. Stasiak,
  Biophys. Jour. {\bf 74}, 2815 (1998).
\bibitem{cieplak} J. I. Su{\l}kowksa, P. Su{\l}kowksa, P. Szymczak and
M. Cieplak, Phys. Rev. Lett. {\bf 100}, 058106 (2008).
\bibitem{liu} Z. Liu, E. L. Zechiedrich, and H. S. Chan, 
  Biophys. J. {\bf 90}, 2344 (2006). 
\bibitem{wasserman} S. A. Wasserman and N. R. Cozzarelli,
Science {\bf 232}, 951 (1986).
\bibitem{sumners} D. W. Sumners,  “Knot theory and DNA,” in New Scientific Applications
of Geometry and Topology, edited by  
D.
W. Sumners, Proceedings of Symposia in Applied Mathematics, Vol. 45,
͑American Mathematical Society, Providence, RI, 1992, 39. 
\bibitem{vologodski} A. V. Vologodski, ̆ A. V. Lukashin, M. D. Frank-Kamenetski ̆ and V. V. Anshelevich, Zh.
Eksp. Teor. Fiz. {\bf 66}, 2153 (1974); Sov. Phys. JETP 39, 1059 (1975);
M. D. Frank-Kamenetskii, A. V. Lukashin and A. V. Vologodskii, Nature
(London) {\bf 258},
398 (1975).
\bibitem{orlandini} E. Orlandini, S. G. Whittington, Rev. Mod. Phys. {\bf 79},
611 (2007);
 C. Micheletti, D. Marenduzzo, and E. Orlandini, Phys. Reports {\bf 504}, 1 (2011).
\bibitem{kurt} T. Vettorel, A. Yu. Grosberg and K. Kremer,
  Phys. Biol. {\bf 6}, 025013  (2009).
\bibitem{mehran} P. Virnau, Y. Kantor and M. Kardar, J. Am. Chem. Soc. {\bf 127} (43),  15102 (2005).
\bibitem{pp2} P. Piera\'nski, S. Przyby{\l} and A. Stasiak, EPJ E
{\bf 6} (2), 123  (2001). 
\bibitem{metzler} R. Metzler, A. Hanke, P. G. Dommersnes, Y. Kantor
  and M. Kardar,
Phys. Rev. Lett. {\bf 88}, 188101 (2002).
\bibitem{muthu} K. Koniaris and M. Muthukumar, Phys. Rev. Lett. {\bf
  66}, 2211 (1991).
\bibitem{YZFF} Y. Zhao and F. Ferrari, J. Stat. Mech. (2012) P11022.
\bibitem{swetnametal} A. Swetnam, C. Brett and M. P. Allen, Phys. Rev. E
  {\bf 85}, 031804 (2012).
\bibitem{scharein} R. Scharein et. al., J. Phys. A: Math. Gen. {\bf
  43}, 475006 (2009).
\bibitem{pivot} N. Madras, A. Orlistsky and L. A. Shepp,  Journal of
  Statistical Physics {\bf 58}, 159 (1990). 
\bibitem{wanglaudau} F. Wang and D. P. Landau, Phys. Rev. Lett. {\bf 86}, 2050 (2001).
\bibitem{pnring} N. A. Volkov, A. A. Yurchenko, A. P. Lyubartsev and
  P. N. Vorontsov-Velyaminov,  Macromol. Theory Simul. {\bf 14},
  491-504 (2005).
\bibitem{pngyrrad} P. N. Vorontsov-Velyaminov, N. A. Volkov,
  A. A. Yurchenko and A. P. Lyubartsev, Polymer Science, Ser. A  {\bf
    52}, 742 (2010).  





\end{thebibliography}
\end{document}